\documentclass[final,3p,times]{elsarticle}
  \usepackage{amssymb}
  \usepackage{amsmath}
  \usepackage{graphics}
  \usepackage{amssymb}
  \usepackage{amsthm}
  \usepackage{setspace}
  \usepackage{epsfig}
  \usepackage{subfigure}
  \usepackage{booktabs}
  \usepackage{epstopdf}
  \biboptions{numbers,sort&compress}
  \usepackage[pdftex,colorlinks,linkcolor=red,anchorcolor=blue,citecolor=green]{hyperref}
  \usepackage{amsmath}
  \usepackage{times}
  \usepackage{anysize}
  \marginsize{2.5cm}{2.5cm}{1cm}{2cm}
  \selectfont
   \journal {XXX}

\begin{document}

      \begin{frontmatter}
   		
   		\newtheorem{theorem}{Theorem}
   		\newtheorem{lemma}[theorem]{Lemma}
   		\title{Efficient method for calculating the eigenvalues of the Zakharov-Shabat system}

   		\author[label2]{ Shikun Cui}
   		\author[label2]{ Zhen Wang  \corref{cor1}  }\ead{wangzmath@163.com}
   		\cortext[cor1]{Corresponding author. School of Mathematical Sciences, Dalian University of Technology, Dalian, 116024, China }
   		\address[label2]{School of Mathematical Sciences, Dalian University of Technology, Dalian, 116024, China}

\begin{abstract}
In this paper, a numerical method is proposed to calculate the eigenvalues of the Zakharov-Shabat system based on Chebyshev polynomials.
A mapping in the form of ${\rm tanh}(ax)$ is constructed according to the asymptotic of the potential function for the Zakharov-Shabat eigenvalue problem.
The mapping could distribute Chebyshev nodes very well considering the gradient for the potential function.
Using Chebyshev polynomials,${\rm tanh}(ax)$ mapping and Chebyshev nodes, the Zakharov-Shabat eigenvalue problem is transformed into a matrix eigenvalue problem, and then solved by the $QR$ algorithm.
This method has good convergence for Satsuma-Yajima potential, and the convergence speed is faster than the fourier collocation method. This method is not only suitable for simple potential functions, but also converges quickly for complex Y-shape potential. This method can also be further extended to solve other linear eigenvalue problems.
\end{abstract}	
   \end{frontmatter}

\section{Introduction}
The NLS equation is an important integrable equation derived from hydrodynamics, it has been used to describe the propagation of optical solitons, langmuir waves in plasma physics, Bose-Einstein condensation and other physical phenomena\cite{JZ-Manakov,JZ-Gross,JZ-Zakharov,JZ-Agrawal}.
The inverse scattering transformation is an important method for solving integrable equations. The inverse scattering transformation of the NLS equation was proposed by Zakharov and Shabat\cite{Zakharov1}.
The Zakharov-Shabat system is the spatial Lax pair of the nonlinear $\rm Schr\ddot{o}dinger$(NLS) equation
\begin{equation}\label{nls}
  {\rm i}q_t+q_{xx}+2\lambda|q|^{2}q=0, \\
\end{equation}
where the subscripts $x$ and $t$ represent the partial derivative with respect to space and time respectively.
When $\lambda=1$, equation (\ref{nls}) is called as the focusing NLS equation, and when $\lambda=-1$, equation (\ref{nls}) is the defocusing NLS equationThe Zakharov-Shabat system has the following form
\begin{align}\label{zs_equation}
&\psi_x=\left(
                     \begin{array}{cc}
                       -{\rm i}k & q \\
                       -\lambda\bar{q} & {\rm i}k \\
                     \end{array}
                   \right)\psi,
\end{align}
where $\psi$ is a column vector, $q$ is the potential function defined in Schwartz space, $" \bar{q} "$ represents the complex conjugation of $q$, $\lambda=\pm1$.

The numerical implementation of the inverse scattering transform attracted special attention when the NLS equation soliton solutions were proposed as potential candidates for fiber optical transmission. At present, increasing the accuracy and efficiency of computational methods for solving the direct Zakharov-Shabat system remains an urgent problem in nonlinear optics.
Calculating the eigenvalues of the Zakharov-Shabat system is an important part in the inverse scattering transform. The number of solitons emerged in the initial profile for the NLS equation is determined by the discrete eigenvalues of the Zakharov-Shabat system.
In most cases, the eigenvalues of the Zakharov-Shabat system (\ref{zs_equation}) cannot be obtained analytically.
It is necessary to develop simple and effective methods for calculating the eigenvalues of the Zakharov-Shabat system.

Up to now, there are some numerical methods were proposed to calculate the eigenvalues of the Zakharov-Shabat system.
Boffetta and Osborne developed a numerical algorithm for computing the direct scattering transform for the NLS equation\cite{Boffetta1992}.
Bronski considered the semi-classical limit of the Zakharov-Shabat eigenvalue problem\cite{Bronski1996}.
The finite difference method was used to compute the Zakharov-Shabat eigenvalue problem numerically\cite{Burtsev1998,Medvedev}.
Hill's method can be used to calculate the eigenvalues of the Zakharov-Shabat system\cite{Deconinck2006,Trogdon2013}.
The Fourier collocation method(FCM) was an effective method to calculate the eigenvalues of the Zakharov-Shabat system\cite{Boyd,yjk}.
Vasylchenkova et al. summarized several Nonlinear Fourier transform(NFT) methods and compare their quality and performance\cite{VPSC2019}.

Above methods can be divided into two types: one is the iterative method for the zero point of Jost function, and the other is to solve the matrix eigenvalue problem\cite{Yousefi}.
Our numerical method belongs to the second type.
We use Chebyshev polynomials and ${\rm tanh}(ax)$ mapping to extract the key information of the potential function, and then transform the Zakharov-Shabat eigenvalue problem into a matrix eigenvalue problem.

The summary of this paper is as follows.  In section \ref{sec_method}, the theoretical knowledge of Chebyshev polynomials is presented and our numerical method is presented in detail. In section \ref{sec_numerical}, the method is used to calculate the eigenvalues of the Zakharov-Shabat system with the Satsuma-Yajima potential, the ${\rm sech}(2\epsilon x){\rm e}^{{\rm i}{\rm sech}(2\epsilon x)/\epsilon}$ potential and the ${\rm exp}(-{\rm i}x){\rm sech}(x)$ potential. The convergence of our method is analyzed. Our method has spectral accuracy, and its convergence rate is fast. Finally, some discussions are given in section \ref{sec_discussion}.
\section{Methodology}\label{sec_method}
In this section, details of our method are introduced.
Our method is summarized as following steps.
For the Zakharov-Shabat system (\ref{zs_equation}),  Chebyshev polynomials are used to approximate the eigenfunction $\psi$ and the potential function $q$ with the help of mapping $H(x)={\rm tanh}(ax) (a>0)$. Using Chebyshev nodes, we turn the Zakharov-Shabat eigenvalue problem into a matrix eigenvalue problem. The $QR$ algorithm is used to calculate the matrix eigenvalue problem, then we can obtain the eigenvalues of the Zakharov-Shabat system.

Defining the $n$ Chebyshev nodes by $$\vec{\chi}=\bigg(-1,\ {\rm cos}(\frac{n-2}{n-1}\pi),\  \cdots \, \ {\rm cos}(\frac{1}{n-1}\pi), 1 \bigg)^{\top}.$$

For the given function $f(x)$ defined in unit interval $\mathbb{I}$, we can  approximate $f(x)$ by its values at $\vec{\chi}$,
\begin{equation}\label{cheb1}
f(x)=T(x)\mathcal{F}f(\vec{\chi}),
\end{equation}
where $T(x)=[T_0(x), \cdots , T_{n-1}(x) ]$, $\mathcal{F}=T(\vec{\chi})^{-1}, f(\vec{\chi})=\Big(\ f(-1), f({\rm cos}(\frac{n-2}{n-1}\pi)), \cdots,f(1)\ \Big)^T$.
$T_k(x) (k=0,1,\cdots,n-1)$ is the Chebyshev polynomial of the first kind,

\begin{equation}\label{nls1}
  \begin{array}{cc}
T_0(x)=1, \nonumber \\
T_1(x)=x, \nonumber \\
 \cdots \\
T_{k}(x)=2xT_{k-1}(x)-T_{k-2}(x). \nonumber
  \end{array}
\end{equation}

Chebyshev polynomials and their derivatives satisfy the relationship\cite{Sezer},
\begin{equation}\label{cheb2}
\frac{\partial}{\partial x}[T_0(x), T_1(x), \cdots , T_{n-1}(x)]=[T_0(x), T_1(x), \cdots , T_{n-1}(x)]\cdot\mathcal{D},
\end{equation}
where $$\mathcal{D}=\left(
                           \begin{array}{cccccc}
                             0 & 1 & 0 & 3 & \cdots & n-1 \\
                             0 & 0 & 4 & 0 & \cdots & 0 \\
                             0 & 0 & 0 & 6 & \cdots & 2(n-1) \\
                             \vdots & \vdots & \vdots & \vdots & \ddots & \vdots \\
                             0 & 0 & 0 & 0 & \cdots & 2(n-1) \\
                             0 & 0 & 0 & 0 & \cdots & 0 \\
                           \end{array}
                         \right)_{n\times n} {\rm for\ odd}\ n,
$$
$$\mathcal{D}=\left(
                           \begin{array}{ccccccc}
                             0 & 1 & 0 & 3 &\cdots& 0 \\
                             0 & 0 & 4 & 0 & \cdots & 2(n-1) \\
                             0 & 0 & 0 & 6 &\cdots & 0 \\
                             \vdots & \vdots & \vdots & \vdots & \ddots& \vdots \\
                             0 & 0 & 0 & 0 & \cdots  & 2(n-1) \\
                             0 & 0 & 0 & 0 & \cdots &0 \\
                           \end{array}
                         \right)_{n\times n} {\rm for\ even}\ n.
$$

Using equation (\ref{cheb1}) and (\ref{cheb2}), the function $\frac{\partial f}{\partial x}$  can be approximated by Chebyshev polynomials,
\begin{equation}\label{cheb3}
\frac{\partial f}{\partial x}=T(x)\mathcal{D}\mathcal{F}f(\vec{\chi}), {\rm for} x\in\mathbb{I}.
\end{equation}
The theoretical knowledge of Chebyshev polynomials has been introduced.

\begin{figure}[h]
  \centering
\vspace{-0.2cm}
  \includegraphics[width=2.4in,height=1.4in]{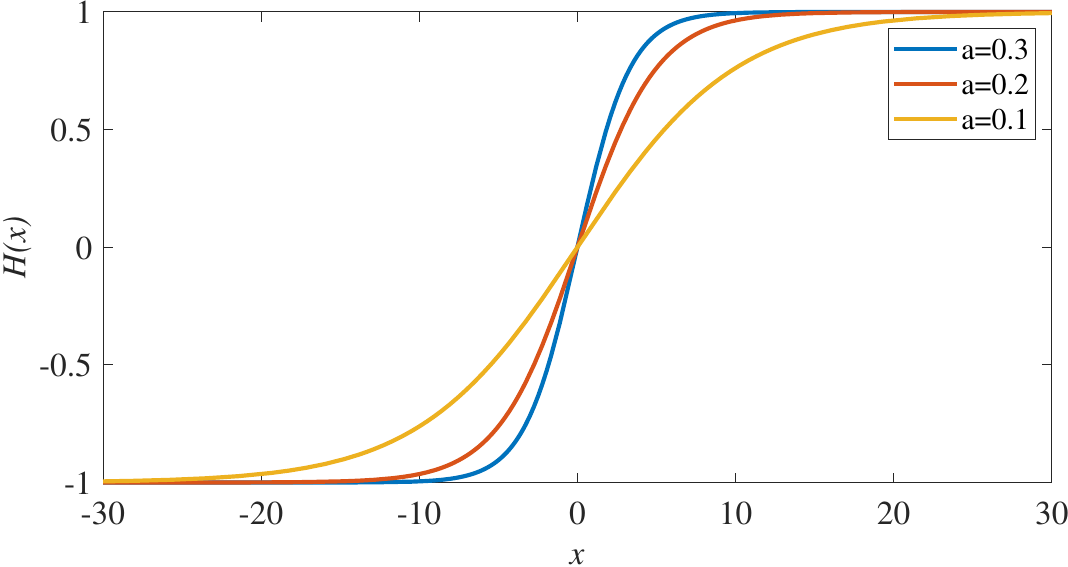}
\vspace{-0.3cm}
  \caption{Results of mapping $H(x)$ about different  $a$.}\label{tanhaax}
\end{figure}

For the given function $g(x)$ defined in real field $\mathbb{R}$, we can approximate $g(x)$ by Chebyshev polynomials and mapping $H(x)={\rm tanh}(ax)$,
\begin{equation}\label{cheb11}
 g(x)=T^{\mathbb{R}}(x)\mathcal{F}g(H^{-1}(\vec{\chi})), x\in\mathbb{R},
\end{equation}
where $T^{\mathbb{R}}(x)=T( H(x) )=[T_0( H(x) ), \cdots, T_{n-1}( H(x) )]$, $H^{-1}$ represent the inverse mapping of $H(x)$.
$H(x)$ is a one-to-one mapping, which maps the real field $\mathbb{R}$ to the unit interval $\mathbb{I}$.
Results of mapping $H(x)$ about different $a$ are shown in Figure \ref{tanhaax}.

Using equations (\ref{cheb2}) and (\ref{cheb11}) and chain rule, $\frac{\partial g}{\partial x}$ can be approximated by Chebyshev polynomials,
\begin{equation}\label{cheb12}
\frac{\partial g(x)}{\partial x}=\frac{\partial H(x)}{\partial x}T^{\mathbb{R}}(x)\mathcal{D}\mathcal{F}g(H^{-1}(\vec{\chi})), {\rm for} x\in\mathbb{R}.
\end{equation}
In this way, the function $g(x)$ and its derivatives $\frac{\partial g}{\partial x}$ are  approximated by Chebyshev polynomials.

If a given function changes rapidly in a certain region, we call this interval as its `rapid-changed interval'.
$H(x)={\rm tanh}(ax)$ changes near 0 rapidly, and its `rapid-changed interval' is expressed as $[-L_1, L_1]$.
$L_1$ is obtained by solving the equation tanh(ax)=$a_1$, where $a_1$ is a real number close to 1.
Taking $a_1=0.9951$ as an example, the `rapid-changed interval' of $H(x)={\rm tanh}(0.3x)$ is $[-10, 10]$, the `rapid-changed interval' of $H(x)={\rm tanh}(0.2x)$ is $[-15, 15]$, the `rapid-changed interval' of $H(x)={\rm tanh}(0.1x)$ is $[-30, 30]$.

The mapping $H(x)={\rm tanh}(0.1x)$  distributes more Chebyshev nodes in the `rapid-changed interval', and distributes less Chebyshev nodes outside the `rapid-changed interval'. So in the `rapid-changed interval', we can effectively identify the key information of the given function with the help of tanh(ax) mapping.

It is worth noting that the value of $a$ will influence the approximate result.
Choosing appropriate $a$ is important in our numerical method.
For the selection of parameter $a$ $(0<a<1)$, we give the following recommendation. The value of $a$ affects the range of `rapid-changed interval', the range of `rapid-changed interval' will increase as $a$ decreases.
For the potential function defined in Schwartz space, it also has the `rapid-changed interval'.
The `rapid-changed interval' of the potential function must be included in the `rapid-changed interval' of tanh($ax$) mapping.
If not, we will not be able to extract the information of the potential function completely.

Rewriting the Zakharov-Shabat system $(\lambda=1)$ into a linear eigenvalue problem
\begin{equation}\label{re_zb}
\left(
  \begin{array}{cc}
    -\frac{\partial}{\partial_x} &q \\
    \bar{q} &\frac{\partial}{\partial_x} \\
  \end{array}
\right)\left(
         \begin{array}{c}
           \psi_1 \\
           \psi_2 \\
         \end{array}
       \right)
={\rm i}k\left(
         \begin{array}{c}
           \psi_1 \\
           \psi_2 \\
         \end{array}
       \right).
\end{equation}

Using equation (\ref{cheb11}) and equation (\ref{cheb12}), we appropriate the eigenfunction $\psi$, $\psi_x$ and the potential function $q$ by Chebyshev polynomials with $n$ nodes,
\begin{equation}\label{sada}
\psi_j=T^{\mathbb{R}}(x)\mathcal{F}\psi_j(H^{-1}(\vec{\chi})),\ \frac{\partial \psi_j}{\partial x}=\frac{\partial H(x)}{\partial x}T^{\mathbb{R}}(x)\mathcal{D}\mathcal{F}\psi_j(H^{-1}(\vec{\chi})),\ q(x)=T^{\mathbb{R}}(x)\mathcal{F}q(H^{-1}(\vec{\chi})),
\end{equation}
where $j=1,2$.

Substituting equation (\ref{sada}) into equation (\ref{re_zb}), we get
\begin{equation}\label{zb_1}
\left(
  \begin{array}{cc}
  -\frac{\partial H(x)}{\partial x}T^{\mathbb{R}}(x)\mathcal{D}\mathcal{F}  &T^{\mathbb{R}}(x)\mathcal{F}q^{\mathbb{R}}(\vec{\chi})T^{\mathbb{R}}(x)\mathcal{F}  \\
  T^{\mathbb{R}}(x)\mathcal{F}\bar{q}^{\mathbb{R}}(\vec{\chi})T^{\mathbb{R}}(x)\mathcal{F}   &\frac{\partial H(x)}{\partial x}T^{\mathbb{R}}(x)\mathcal{D}\mathcal{F}  \\
  \end{array}
\right)\left(
         \begin{array}{c}
          \psi_1^{\mathbb{R}}(\vec{\chi}) \\
          \psi_2^{\mathbb{R}}(\vec{\chi}) \\
         \end{array}
       \right)={\rm i}k\left(
         \begin{array}{c}
          T^{\mathbb{R}}(x)\mathcal{F}\psi_1^{\mathbb{R}}(\vec{\chi}) \\
          T^{\mathbb{R}}(x)\mathcal{F}\psi_2^{\mathbb{R}}(\vec{\chi}) \\
         \end{array}
       \right),
\end{equation}
where $q^{\mathbb{R}}(\vec{\chi})=q(H^{-1}(\vec{\chi}))$, $\bar{q}^{\mathbb{R}}(\vec{\chi})=\bar{q}(H^{-1}(\vec{\chi}))$, $\psi_j^{\mathbb{R}}(\vec{\chi})=\psi_j(H^{-1}(\vec{\chi})) (j=1,2)$.

Setting $x$=$H^{-1}(\vec{\chi})$, equation (\ref{zb_1}) is rewritten into
\begin{equation}\label{zb_2}
\left(
  \begin{array}{cc}
  -{\rm diag}\Big[\frac{\partial H(\vec{\chi})}{\partial x} \Big]\mathcal{F}^{-1}\mathcal{D}\mathcal{F}
&{\rm diag}\Big[q(H^{-1}(\vec{\chi}))\Big]  \\
  {\rm diag}[\bar{q}(H^{-1}(\vec{\chi}))]   &{\rm diag}\Big[\frac{\partial H(\vec{\chi})}{\partial x} \Big]\mathcal{F}^{-1}\mathcal{D}\mathcal{F}  \\
  \end{array}
\right)_{2n\times2n}\left(
         \begin{array}{c}
          \psi_1^{\mathbb{R}}(\chi) \\
          \psi_2^{\mathbb{R}}(\chi) \\
         \end{array}
       \right)_{2n\times1}={\rm i}k\left(
         \begin{array}{c}
         \psi_1^{\mathbb{R}}(\chi) \\
         \psi_2^{\mathbb{R}}(\chi) \\
         \end{array}
       \right)_{2n\times1},
\end{equation}
where$${\rm diag}\Big[\frac{\partial H(\vec{\chi})}{\partial x} \Big]=\left(
                                                                        \begin{array}{cccc}
                                                                          \frac{\partial H(x)}{\partial x}\mid_{x=-1} & 0 & \cdots & 0 \\
                                                                          0&\frac{\partial H(x)}{\partial x}\mid_{x={\rm cos}(\frac{n-2}{n-1}\pi)}  &\cdots  &0  \\
                                                                           \vdots&\vdots  & \ddots & \vdots \\
                                                                          0 &0 & \cdots &\frac{\partial H(x)}{\partial x}\mid_{x=1}  \\
                                                                        \end{array}
                                                                      \right).
$$

Equation (\ref{zb_2}) is recorded as $A\psi={\rm i}k\psi$, where $$A=\left(
                                                                    \begin{array}{cc}
                                                                      -A_1 & B \\
                                                                      B^{\pounds} &  A_1 \\
                                                                    \end{array}
                                                                  \right),
$$
where $A_1={\rm diag}\Big[\frac{\partial H(\vec{\chi})}{\partial x} \Big]\mathcal{F}^{-1}\mathcal{D}\mathcal{F}$,
$B^{\pounds}$ is the Hermitian of $B$, $B={\rm diag}\Big[q(H^{-1}(\vec{\chi}))\Big]$.
In fact, equation (\ref{zb_2}) is a $2n\times2n$ matrix eigenvalue problem.
Note that $A_1$ is the differentiation matrix for $\partial_x$ in our method, $B$ is a diagonal matrix composed of Chebyshev series for potential $q(x)$.

The eigenvalue problem (\ref{zb_2}) can be solved by the $QR$ algorithm\cite{Parlett2000}.
In the $QR$ algorithm, the matrix $A$ is decomposed into $A=QR$, where $Q$ is a orthogonal matrix and $R$ is an upper triangular matrix.

Steps of the $QR$ algorithm is as follows.
\begin{align}\label{111}
&A_1=A=Q_1R_1, \nonumber\\
&A_2=R_1Q_1=Q_1^{\top}A_1Q_1=Q_2R_2, \nonumber \\
&\cdots \nonumber\\
&A_n=Q_nR_n=(Q_1Q_2...Q_k)^{\top}A(Q_1Q_2...Q_k), \nonumber
\end{align}
diagonal elements of $A_n$ are the eigenvalues of $A$ as $n\rightarrow\infty$.

Regarding the accuracy of this method, the method do not need to truncate the interval, and the method has spectral accuracy\cite{Boyd} for smooth potential function.
Because we do not truncate the calculated interval, our method will not produce truncation error for analytic potential.
\section{Numerical results}\label{sec_numerical}
Our method is used to calculate the eigenvalues of Zakharov-Shabat system($\lambda =1$) (\ref{zs_equation}) with three potentials, and the convergency of the method is analysed.
All numerical examples reported here are run on a Asustek computer with Intel(R) Core(TM) i7-11800H processor and 16 GB memory.
\subsection{Satsuma-Yajima potential function.}
Our numerical method is used to calculate the eigenvalues of the Zakharov-Shabat system($\lambda =1$) (\ref{zs_equation}) with Satsuma-Yajima potential $A{\rm sech}(x)$.
Numerical results are compared with the analytical results, and the performance of our numerical method is compared with the performance of the FCM\cite{yjk}.

When $q=A{\rm sech}(x)$ , Satsuma and Yajima exactly calculated the discrete eigenvalues of the Zakharov-Shabat system\cite{SATSUMA1974}.
Satsuma and Yajima found the discrete eigenvalue in upper half complex plane $\mathbb{C}_+$ is
\begin{equation}\label{dis}
\kappa_n={{\rm i}\Big(A+\frac{1}{2}-n \Big) },
\end{equation}
where $n$ is a positive number satisfying $n\leq A+\frac{1}{2}$. Due to symmetry of the discrete eigenvalues\cite{SATSUMA1974}, the Zakharov-Shabat system with $A{\rm sech}(x)$ potential has the discrete eigenvalues $\bar{\kappa}_n$ in $\mathbb{C}_{-}$.

\begin{figure}[h]
  \centering
  \subfigure[]{\includegraphics[width=1.8in,height=1.66in]{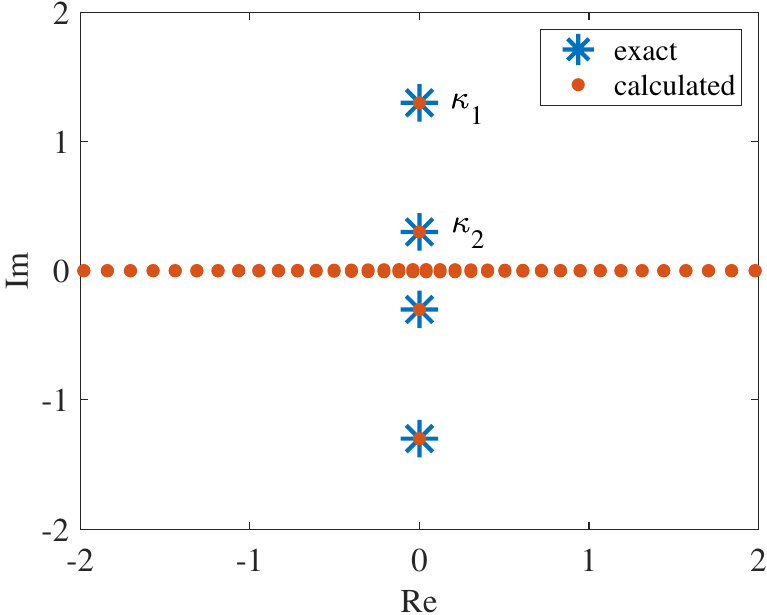}}
  \subfigure[]{\includegraphics[width=1.7in,height=1.66in]{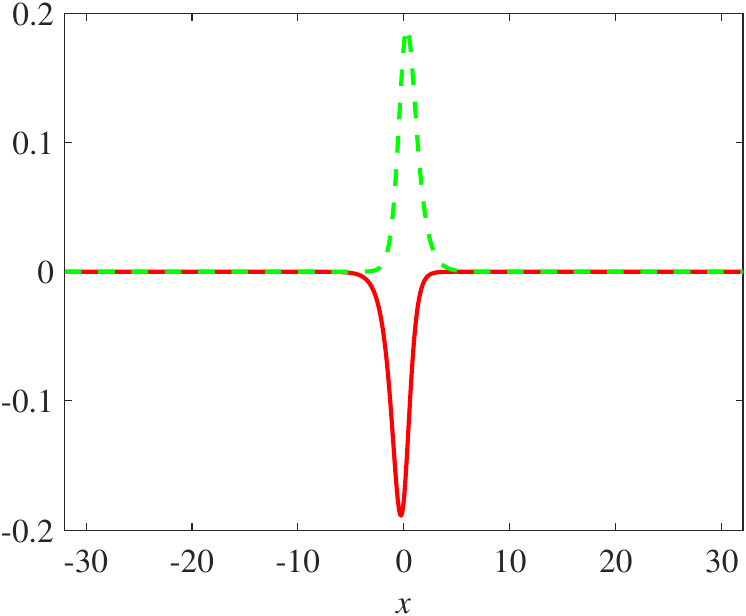}}
  \subfigure[]{\includegraphics[width=1.7in,height=1.66in]{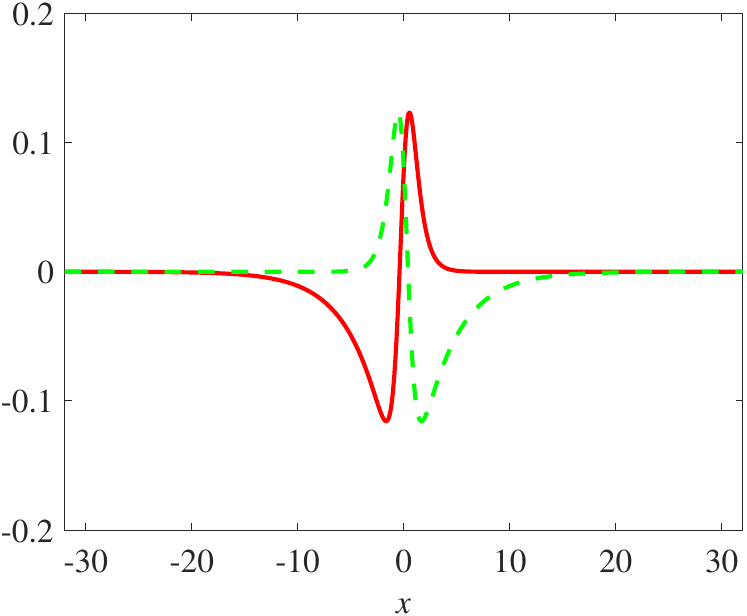}}
  \caption{The calculated results of Zakharov-Shabat system with $1.8{\rm sech}(x)$ potential. (a): The calculated eigenvalues(red) and exact eigenvalues(blue) of $1.8{\rm sech}(x)$ potential. (b): Numerical results of calculated eigenfunctions $\psi_1$(red line) and $\psi_2$(green line) when $k=\kappa_1$. (c): Numerical results of calculated eigenfunctions $\psi_1$(red line) and  $\psi_2$(green line) when $k=\kappa_2$.}
\label{result_11}
\end{figure}

In the specific calculation, we calculate the eigenvalues of $q(x)=1.8{\rm sech}(x)$.
When $q(x)=1.8{\rm sech}(x)$, the Zakharov-Shabat system has four discrete eigenvalues $\kappa_1=1.3{\rm i}$, $\kappa_2=0.3{\rm i}$, $\bar{\kappa}_1=-1.3{\rm i}$ and $\bar{\kappa}_2=-0.3{\rm i}$.
The number of Chebyshev nodes $n$  is set to 200, the value of $a$ is set to 0.15.
The calculating results are shown in Figure \ref{result_11}.
Figure \ref{result_11}(a) shows the calculated eigenvalues of the Zakharov-Shabat system with the Satsuma-Yajima potential.
There are four discrete eigenvalues in Figure \ref{result_11}(a), which is consist with the theoretical result.
Figure \ref{result_11}(b) shows the calculated eigenfunction in point $\kappa_1$, and Figure \ref{result_11}(c) gives the calculated eigenfunction in point $\kappa_2$. The absolute error between the calculated $\kappa_1$ and the exact $\kappa_1$ is $1.85\times10^{-15}$, and the absolute error between the calculated $\kappa_2$ and the exact $\kappa_2$ is $1.61\times10^{-16}$. The method takes about 0.25 seconds to finish.
Above results show that our method is efficient.



\begin{figure}[h]\label{error1}
  \centering
  \subfigure[Absolute error.]{\includegraphics[width=2.1in,height=1.7in]{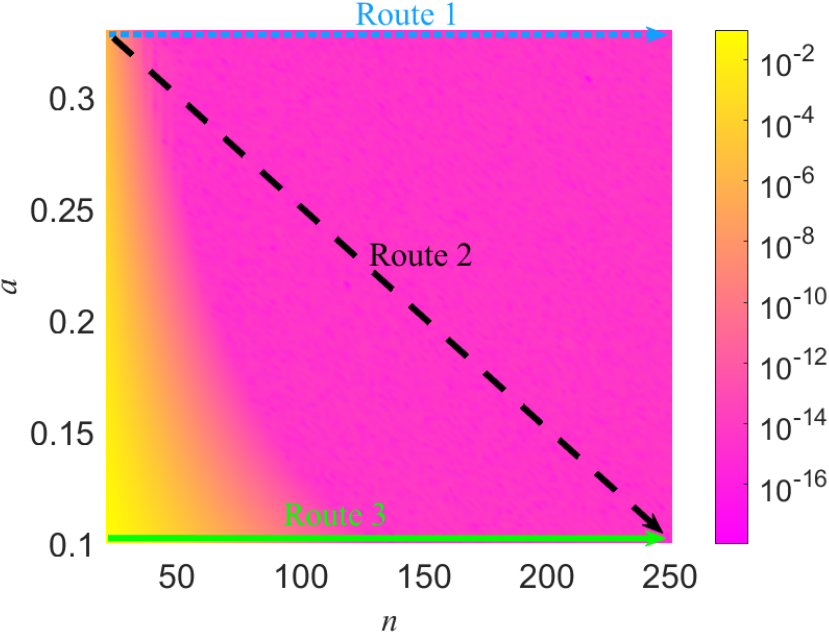}}
  \subfigure[Error diagram versus the number of nodes.]{\includegraphics[width=2.5in,height=1.67in]{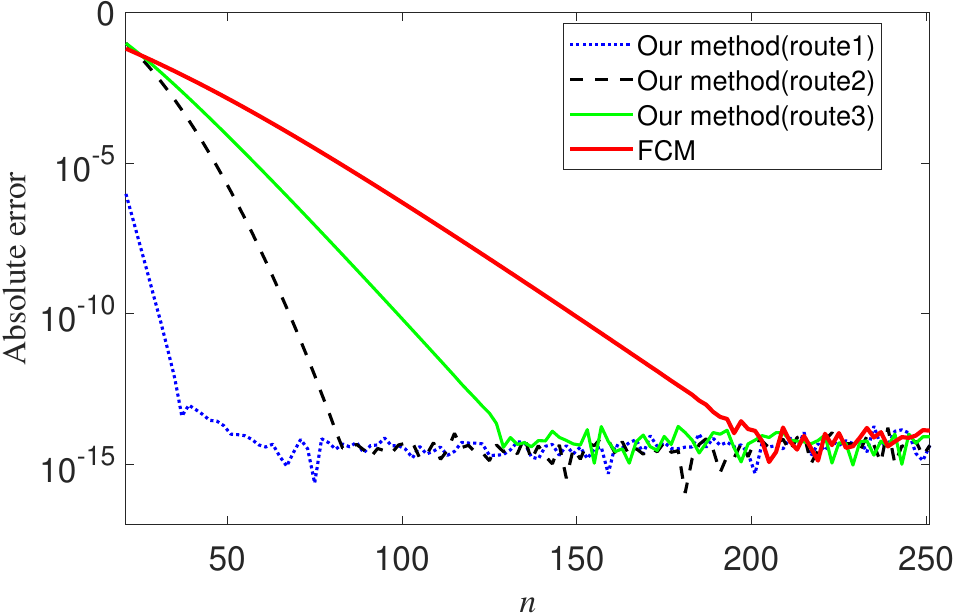}}
  \caption{The calculated absolute error in area $[a,n]\in[0.1,0.33]\times[21,251]$. Subfigure(a): the absolute error picture($k=\kappa_1$). Subfigure(b): the error diagram along Route 1 (blue line), the error diagram along Route 2 (black line), the error diagram along Route 3 (green line) and error diagram of the Fourier collocation method (red line). }
\end{figure}

The stability and convergency of our method needs to be analyzed.
In area $[a,n]\in[0.1,0.33]\times[21,251]$, we calculate the eigenvalues of the Zakharov-Shabat system with $q(x)=1.8{\rm sech}(x)$, and the absolute error in $k=\kappa_1$ is shown in Figure \ref{error1}(a).
There are three routes in Figure \ref{error1}(a) (blue Route 1, black Route 2, and green Route 3), the convergency of our method is analyzed along the three routes.
In the Fourier collocation method, the calculated interval is truncated to [-25, 25].
The relationship between the error and the number of $n$ nodes is shown in Figure \ref{error1}(b), the red line is the error curve calculated by the Fourier collocation method(FCM), the blue line is the error curve calculated by our method along Figure \ref{error1}(a) ``Route 1'', the black line is the error bar calculated by our method along Figure \ref{error1}(a) ``Route 2'', and the green line is the error bar calculated by our method along Figure \ref{error1}(a) ``Route 3''.
Figure \ref{error1}(b) shows that our method is more accurate than the FCM, and the convergence rate of our method is faster than FCM, so our method is more efficient.
Because the error calculated by the FCM decays exponentially with the number of nodes\cite{yjk}, the error of our method also decays exponentially with the number of nodes, its error decays faster than any power of $n^{-1}$. Thus spectral accuracy of the method is confirmed.

The minimum error generated by our method is about $10^{-15}$ level.
The error is caused by the calculation accuracy of the software.
Since the calculation accuracy of the mathematical software is 16 significant figures, there will be an error of about $10^{-15}$ level in the calculation process. Our method can greatly improve the calculation accuracy, especially when the number of Chebyshev nodes is small.

\subsection{Y-shape potential}
\begin{figure}[h]
  \centering
  \subfigure[$\epsilon=0.2$]{\includegraphics[width=1.7in,height=1.5in]{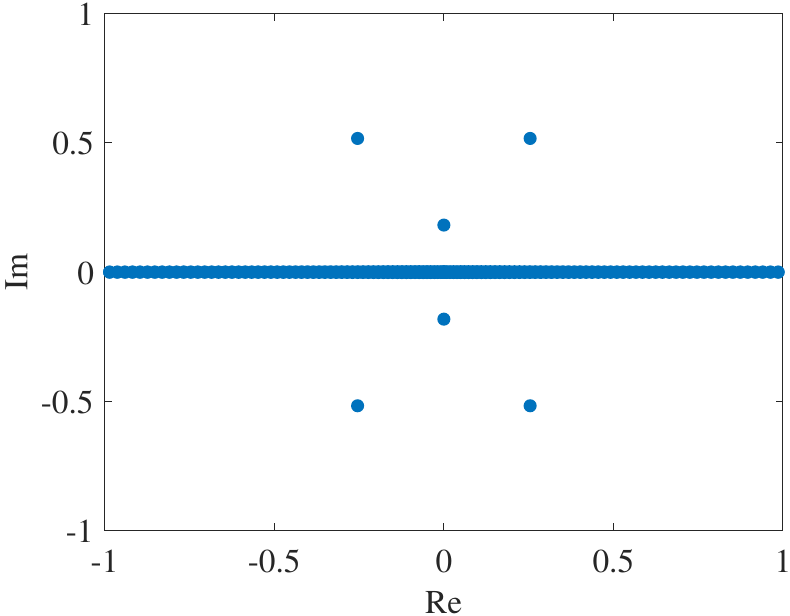}}
  \subfigure[$\epsilon=0.1$]{\includegraphics[width=1.7in,height=1.5in]{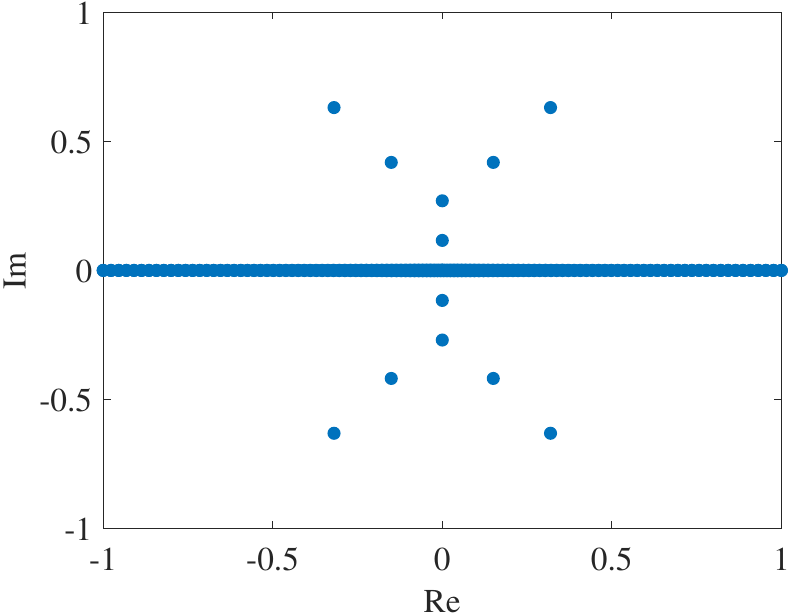}}
  \subfigure[$\epsilon=0.05$]{\includegraphics[width=1.7in,height=1.5in]{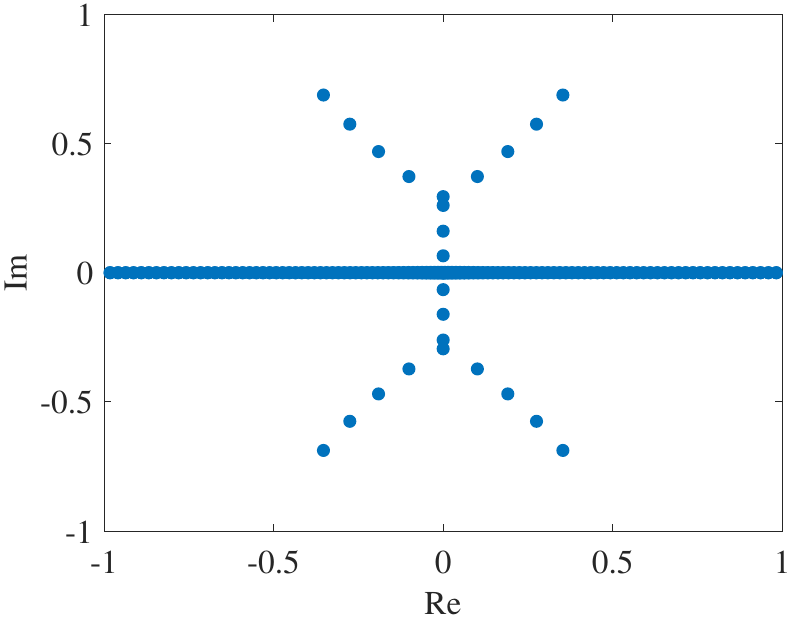}}
  \caption{The calculated eigenvalues of Zakharov-Shabat system with ${\rm sech}(2\epsilon x){\rm e}^{{\rm i}{\rm sech}(2\epsilon x)/\epsilon}$ potential(a=0.02).}\label{ex1}
\end{figure}

Bronski computed the eigenvalues of the ${\rm sech}(2\epsilon x){\rm e}^{{\rm i}{\rm sech}(2\epsilon x)/\epsilon}$ potential and found the shape of the discrete eigenvalues is ``Y''\cite{Bronski1996}. Setting $n=400$ and $a=0.02$, our method is used to compute the eigenvalues of the ${\rm sech}(2\epsilon x){\rm e}^{{\rm i}{\rm sech}(2\epsilon x)/\epsilon}$ potential with $\epsilon=0.2$, $\epsilon=0.1$ and $\epsilon=0.05$, and the calculated results are shown in Figure \ref{ex1} respectively.
The calculations are finished within 0.6 seconds.

From Figure \ref{ex1}, there are three discrete eigenvalues in $\mathbb{C}_{+}$ when $\epsilon=0.2$, six discrete eigenvalues in $\mathbb{C}_{+}$ when $\epsilon=0.1$, and twelve discrete eigenvalues in $\mathbb{C}_{+}$ when $\epsilon=0.05$. The calculated results are consist with Bronski's results(\cite{Bronski1996}, page385, Table 1). The calculated discrete eigenvalues become Y-shape with the decrease of $\epsilon$, which are consist with the theoretical results.

There are six discrete eigenvalues in $\mathbb{C}_{+}$ for Figure \ref{ex1}(b), and their values are shown in Table \ref{eps_1}.
\begin{table}[h]
\centering
\begin{tabular}{|c|r|}
  \hline
  No. & Value \\
\hline
  $\kappa_1$ &  -1.78524894765016e-15 + 0.116148026898534i\\
  $\kappa_2$ &  5.16823894592694e-15 + 0.269496534408172i\\
  $\kappa_3$ &  0.150457991591637 + 0.418161274246707i\\
  $\kappa_4$ &  -0.150457991591641 + 0.418161274246702i\\
  $\kappa_5$ &  0.319248334509386 + 0.630381427554910i\\
  $\kappa_6$ &  -0.319248334509384 + 0.630381427554907i\\
  \hline
\end{tabular}\label{eps_1}
\end{table}

\begin{figure}[h]
  \centering
  \subfigure[2D image]{\includegraphics[width=2in,height=1.6in]{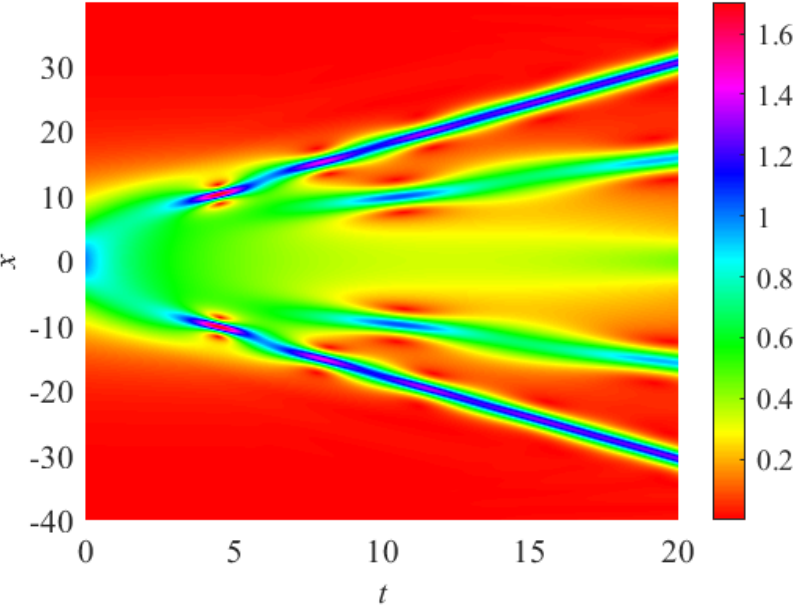}}
  \subfigure[3D image]{\includegraphics[width=2.1in,height=1.8in]{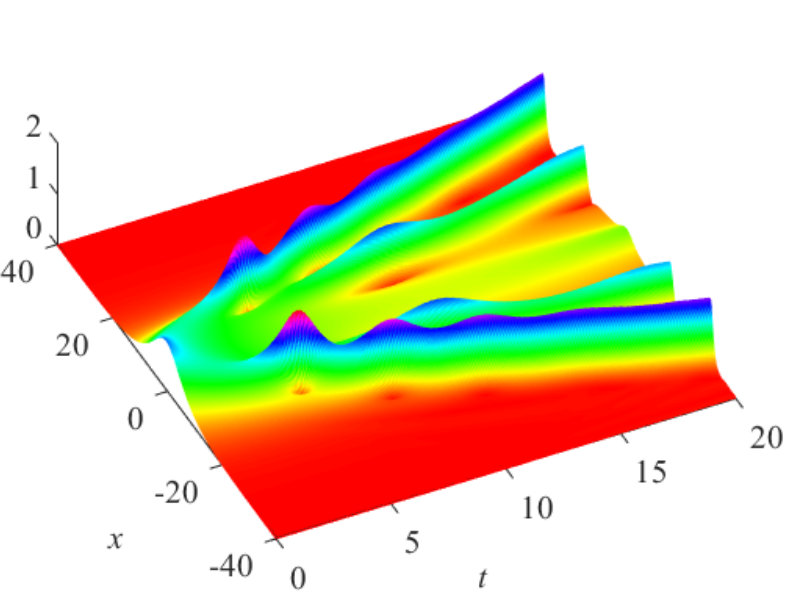}}
  \caption{The evolution of the initial profile $q(x,0)={\rm sech}(0.2x){\rm e}^{{\rm 10i}{\rm sech}(0.2x)}$ for the NLS equation.}\label{fig_ysech}
\end{figure}

From Table \ref{eps_1}, we learn that the ${\rm sech}(2\epsilon x){\rm e}^{{\rm i}{\rm sech}(2\epsilon x)/\epsilon}$ potential has two pure imaginary eigenvalue and four complex discrete eigenvalues.
Thus the ${\rm sech}(2\epsilon x){\rm e}^{{\rm i}{\rm sech}(2\epsilon x)/\epsilon}$ initial profile will evolve into a second-order breather and four solitons for the NLS equation. The fourier spectrum method\cite{JieS} is used to calculate the evolution of the NLS equation with the ${\rm sech}(2\epsilon x){\rm e}^{{\rm i}{\rm sech}(2\epsilon x)/\epsilon}$ initial profile.
The density of the calculated result is shown in Figure \ref{fig_ysech}. In Figure \ref{fig_ysech}(a) and Figure \ref{fig_ysech}(b), the initial profile $q(x,0)={\rm sech}(0.2x){\rm e}^{{\rm 10i}{\rm sech}(0.2x)}$ evolves into four solitons and a second-order breather, which is consist with Figure \ref{ex1}(b).

\begin{figure}[h]
  \centering
  \includegraphics[width=2.2in,height=1.5in]{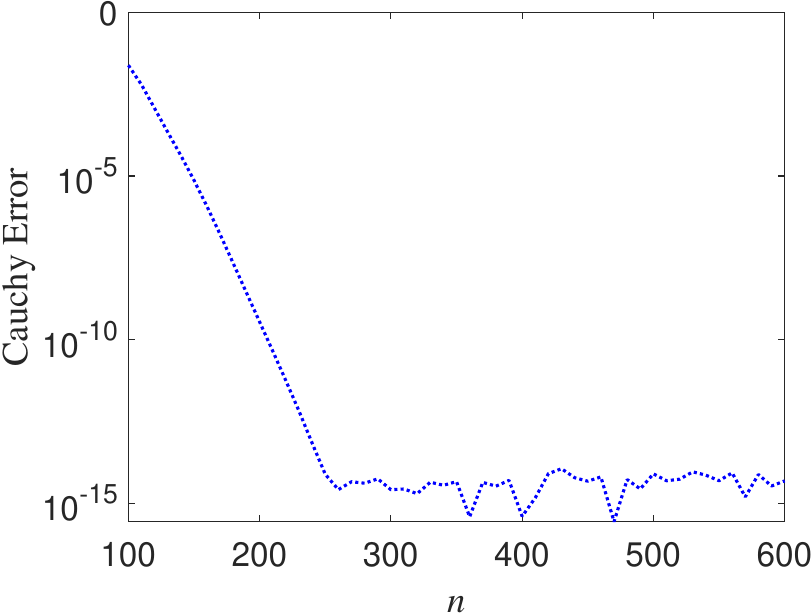}
  \caption{The cauchy error for the ${\rm sech}(2\epsilon x){\rm e}^{{\rm i}{\rm sech}(2\epsilon x)/\epsilon}$ in $\kappa_1$.}\label{fig_conver}
\end{figure}

The correctness of the calculation results is verified by analyzing the convergence of the method.
When $n=400$ and $a=0.02$, we obtain the eigenvalue $\kappa_1= -1.78524894765016\cdot10^{-15} + 0.116148026898534{\rm i}$ of the ${\rm sech}(2\epsilon x){\rm e}^{{\rm i}{\rm sech}(2\epsilon x)/\epsilon}$ potential.
Under different $n$ Chebyshev nodes, we calculate the cauchy error for the ${\rm sech}(2\epsilon x){\rm e}^{{\rm i}{\rm sech}(2\epsilon x)/\epsilon}$ potential in $\kappa_1$. The calculated result is shown in Figure \ref{fig_conver}.
The cauchy error is the absolute error between the calculated result and $\kappa_1= -1.78524894765016\cdot10^{-15} + 0.116148026898534{\rm i}$.
In Figure \ref{fig_conver}, the method gradually converges as $n$ increases, and generates an error of $10^{-15}$ level.
\subsection{solitonic potential}
\begin{figure}[h]
  \centering
  \includegraphics[width=2in,height=1.8in]{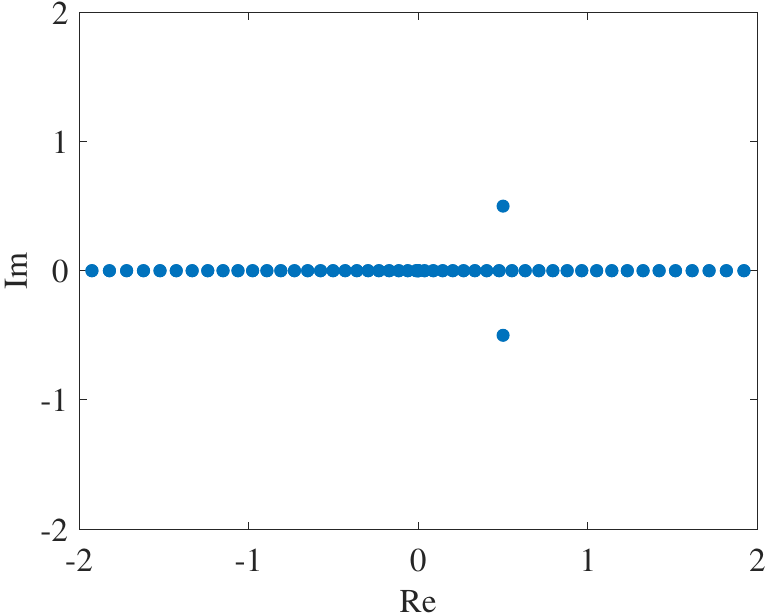}\\
  \caption{The calculated eigenvalues of $q_{so}={\rm exp}(-{\rm i}x){\rm sech}(x)$ potential.}\label{one_so}
\end{figure}

In the end, we also calculate the eigenvalues for the solitonic potential $q_{so}={\rm exp}(-{\rm i}x){\rm sech}(x).$  As we all know, $q_{so}$ has the single discrete eigenvalue $\kappa_1=0.5+0.5{\rm i}$ in $\mathbb{C}_{+}$\cite{Burtsev1998}. Setting $n=200$ and $a=0.1$, our method is used to compute the spectrum of $q_{so}={\rm exp}(-{\rm i}x){\rm sech}(x)$, the calculated result is shown in Figure \ref{one_so}. The absolute between the calculated $\kappa_1$ and the exact $\kappa_1$ is $7.77\times10^{-16}$, the absolute error between the calculated $\kappa_2$ and the actual $\kappa_2$ is $6.31\times10^{-15}$.
The calculation is finished within 0.3 seconds.
Our method is more accurate and faster than the NFT method\cite{VPSC2019}.

\section{Conclusion}\label{sec_discussion}
A numerical algorithm is proposed to solve the Zakharov-Shabat eigenvalue problem. The used tools are Chebyshev polynomials, ${\rm tanh}(ax)$ mapping and the $QR$ algorithm.
We can effectively identify the key information of the given function with
the help of ${\rm tanh}(ax)$ mapping and realize the high-efficiency calculation.

The method has following advantages.
First, we do not need to truncate the calculated region for analytical potentials, so our method will not produce truncation error when using Chebyshev polynomials to appropriate the given function.
Second, the method can calculate the discrete eigenvalues for the Zakharov-Shabat system with spectral accuracy. The method is high-precision and efficient.
We calculate the discrete eigenvalues of the Satsuma-Yajima potential, and compare the method with the Fourier collocation method, the convergence rate of our method is faster than the Fourier collocation method.
For complex ${\rm sech}(2\epsilon x){\rm e}^{{\rm i}{\rm sech}(2\epsilon x)/\epsilon}$ potential, the method still converge quickly.
It is worth mentioning that this method can also be further extended to solve other linear eigenvalue problems.

\section*{Acknowledgment}
This project is supported by NSFC (52171251), LiaoNing Revitalization Talents Program (XLYC1907014) and ``the Fundamental Research Funds for the Central Universities" (DUT21ZD205).


  \end{document}